# Spatial embedding of neuronal trees modeled by diffusive growth

*(to appear in J Neurosc Methods 2006)*

Artur Luczak

Center for Molecular and Behavioral Neuroscience
Rutgers University, Newark, NJ
E-mail: Luczak@rutgers.edu
March 2006

**Abstract**
The relative importance of the intrinsic and extrinsic factors determining the variety of geometric shapes exhibited by dendritic trees remains unclear. This question was addressed by developing a model of the growth of dendritic trees based on diffusion-limited aggregation process. The model reproduces diverse neuronal shapes (i.e., granule cells, Purkinje cells, the basal and apical dendrites of pyramidal cells, and the axonal trees of interneurons) by changing only the size of the growth area, the time span of pruning, and the spatial concentration of 'neurotrophic particles'. Moreover, the presented model shows how competition between neurons can affect the shape of the dendritic trees. The model reveals that the creation of complex (but reproducible) dendrite-like trees does not require precise guidance or an intrinsic plan of the dendrite geometry. Instead, basic environmental factors and the simple rules of diffusive growth adequately account for the spatial embedding of different types of dendrites observed in the cortex. An example demonstrating the broad applicability of the algorithm to model diverse types of tree structures is also presented.

**Key words:** Diffusion-limited aggregation; Neuronal morphology; Dendrites; Growth model; dendritic geometry.

## 1. Introduction

The geometry of dendritic trees plays an important role in determining the connectivity (Amirikian, 2005; Stepanyants and Chklovskii, 2005) and electrophysiological properties of neurons (Migliore et al. 1995; Mainen and Sejnowski, 1996; Krichmar et al., 2002). However, the extent to which intrinsic and extrinsic factors shape dendritic geometry remains largely unknown (Scott and Luo, 2001). The complexity of interactions between different intrinsic and extrinsic factors during the development of neuronal arborization can make it very difficult to separate their contributions experimentally. In this study, I propose a simple computational model that demonstrates that two basic external factors – (i) the space available for growth and (ii) the spatial distribution of 'neurotrophic particles' (NPs) – can adequately account for the three-dimensional (3D) embedding of dendritic and axonal trees.

Although past models of dendritic growth have been proposed, none of these models considered the role of environmental factors on 3D structure of neurons. For example, in some of earlier works, only dendrograms were modeled (i.e., connectivity among branches, and their length and diameter) yet spatial embedding was not considered (Nowakowski et al., 1992; Van Pelt et al., 1997; Van Ooyen et al., 2001). In the 3D models of dendritic trees, several parameters measured from real neurons (e.g., the probability distribution of branching points as a function of the distance from a soma) were used and stochastic procedures were applied to recreate dendrites while disregarding influence of environment (Ascoli, 1999; Burke and Marks 2002; Samsonovich and Ascoli, 2003; Samsonovich and Ascoli, 2005; for review see Ascoli, 2002).

In contrast to the above models based on statistical reconstruction of dendrites, the present model simulates 3D neuronal growth using external factors. In this approach, dendrite geometry parameters (e.g., number of segments, branching probability, orientation, etc.) are not built into the model but rather geometry parameters emerge as a result of environmental factors such as the NP concentration, competition between neurons, and space limitations. External cues are well known to play a significant role in shaping dendritic geometry (Horch and Katz 2002), and hence the model presented here accounts for the important biological processes underlying neuronal geometry (see Discussion).

To simulate neuronal growth, I used diffusion-limited aggregation (DLA), which is a well-established physical model for the formation of structures controlled by diffusion processes (Witten and Sander, 1981). Prior research has demonstrated that DLA can provide a good description of a variety of natural processes, such as electrical discharge in gas (lightning) (Niemeyer et al., 1984), electrochemical deposition (Halsey, 1990; Brady and Ball, 1984), or the growth of snowflakes (Family et al., 1987). The form of a typical DLA structure is illustrated in the insert of Figure 1.

Previously, diffusive processes were invoked to explain the origin of dendritic arbors by Hentschel and Fine (1996), who proposed a two-dimensional diffusion-regulated model of dendritic morphogenesis in which cell growth depended upon the local concentration of calcium. However, their model was restricted to the early stage of neuronal growth only, and was based on *in vitro* cultures. In order to generate a 3D embedding of fully developed dendritic trees, the present model operates at a coarser level. It takes into account the local concentrations of NPs, but without including such details as changes in the concentrations of ions along the dendrite membrane.

In the presented DLA-based model, assuming only that neurons grow in the direction of a local gradient of neurotrophic substance and that dendrites compete for the same resources, it was possible to reproduce the spatial embedding of major types of cerebral neurons: granule cells, Purkinje cells, pyramidal cells, and dendritic and axonal trees of interneurons. Interestingly, the same model can be applied to model other types of tree structures, as shown here using the example of a generated root and two types of real tree, which suggests similarities between the mechanism of dendrite growth and the variety of branched structures, where the objective is to optimize access to tropic factors.

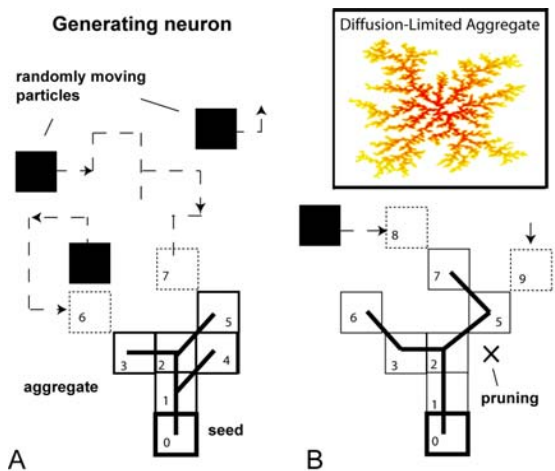

**Figure 1.**
Illustration of the DLA algorithm. (A) Randomly moving particles (black) stick irreversibly at their point of first contact with the aggregate (composed of particles 0–5). To each newly jointed particle a parent particle is assigned and both become connected by a line segment. (B) While the aggregate grows, the particles at the terminals are randomly deleted from the aggregate (pruning) during a specified time window. Insert: Example of a two-dimensional DLA comprising 6000 particles. The color intensity decreases in the order in which particles connected to the aggregate.

## 2. Materials and Methods
### 2.1. Generating neurons

The growth rule for DLA can be defined inductively as follows: introduce a randomly moving particle at a large distance from an $n$-particle aggregate, which sticks irreversibly at its point of first contact with the aggregate, thereby forming an $n + 1$ particle aggregate. Figure 1A illustrates a sample trajectory of particles that stick to an aggregate composed of five particles (each particle is numbered in the order in which it contacts the aggregate; the seed particle is numbered 0). Stated differently, the aggregate grows by one step at the point of contact with a particle, thus prominent branches screen internal regions of the aggregate, preventing them from growing further (Halsey, 1997). For computational efficiency, instead of one moving particle, $m$ simultaneously moving particles were introduced (Voss, 1984). In the presented model, the initial distribution of particles is a model parameter and thus particles are not always uniformly distributed, which is a significant difference from the classical DLA. As a result, there is a higher probability that the aggregate will grow toward a higher concentration of particles.

For computational simplicity, DLA was generated on a 3D square grid inside a rectangular box. At every iteration step, particles moved by one position in the grid according to the Margolus rule (Toffoli and Margolus 1998), which results in a pseudorandom movement of particles and increases computational efficiency. Illustration of the initial spatial distribution of particles for the granule cells is presented in Figure 2A. The number of seed particles placed inside a box determined the number of aggregates. To simulate an ensemble of simultaneously growing neurons, I used nine equally distributed seeds (Fig. 2, Table I).

As the new particles connect to the aggregate, a parent particle is assigned to each newly connected particle at the point of its connection to the aggregate. When a new particle is connected to more than one particle in the aggregate, the parent particle is selected at random. For example, in Figure 1A, for particle number 4, either particle number 1 or particle number 2 could be assigned as a parent particle, and in this case particle 1 was selected at random. Thus the aggregate is converted to a directed, acyclic tree, where each particle becomes a node connected by a segment to an assigned parent node. In a 3D grid, a particle can contact up to 26 neighboring particles (later in the text the particles are also referred to as NPs).

Without additional restrictions, the aggregate would form a heavily branched structure similar to DLA in the insert of Figure 1. Therefore, I implemented a pruning procedure, which removes terminal particles from the aggregate. At each iteration there is probability $p = 0.4$ that any terminal particle of the aggregate can be deleted if that particle was connected within the last PS iterations, but later than 5 iterations ago, where PS is a pruning span parameter. As a result of the deletion, the parent particle of the removed particle becomes again a terminal particle (eligible for the deletion) unless it is a branching node. Thus increasing pruning span increases the number of deleted particles. Five iterations were chosen before applying pruning, primary to allow for the initial growth of new branches. Nevertheless, this parameter has a very minor effect on the geometry of a dendrite as compared to pruning span. The removed particles do not return to the pool of NPs and the seed particle cannot be removed by definition. The algorithm stops when no new particle is connected for 100 iterations.

The resulting structure was smoothed to reduce the regularity artifact introduced by the use of a uniform grid. Additionally, the use of a grid increased the tri- and higher order furcations, because segments could connect to a tree only at discrete points. To correct this artifact all furcations were reduced to bifurcations by splitting a node with $x$ segments into $x–2$ randomly connected nodes shifted by a small, random amount from the original location. Generating DLA with an out-of-grid algorithm would prevent the occurrence of the above artifacts, but this would increase the computational time by at least an order of magnitude. Another solution to this problem could be to substantially increase the growing space, which would decrease the relative size of NPs and thereby reduce the regularity artifact. Unfortunately, the computational time would increase exponentially. For instance, for a box with $30\times30\times30$ particles it takes 51 s to create DLA and for a box with $60\times60\times60$ particles the corresponding time is 2906 s (both cases using the following parameters: particle density = 0.5, number of aggregates = 1; PS = 20; MATLAB 7.1, PC with 2.4-GHz Intel processor and 2 GB of RAM). I estimate that to model a slab of tissue with a realistic density of cells would require particles at least 20 times smaller that those used here, which would make the computations unacceptably long.

The parameters used to generate different types of neurons were optimized manually and are specified in Table I. The MATLAB code used to produce the described simulations is available at http://bin.yale.edu/~ajl37/artur.html or upon request.

### 2.2. Branch diameter

The aim of the present model is to explain the 3D embedding of neurons, and hence the branch diameter lies outside the scope of this work. Nevertheless, a diameter can be easily assigned to each branch of a generated dendrite based on the distance from the dendrite tip. For instance, Samsonovich and Ascoli (2005) presented that the branch diameter of hippocampal pyramidal cells can be approximated by a linear function of the topological distance from terminal segments.

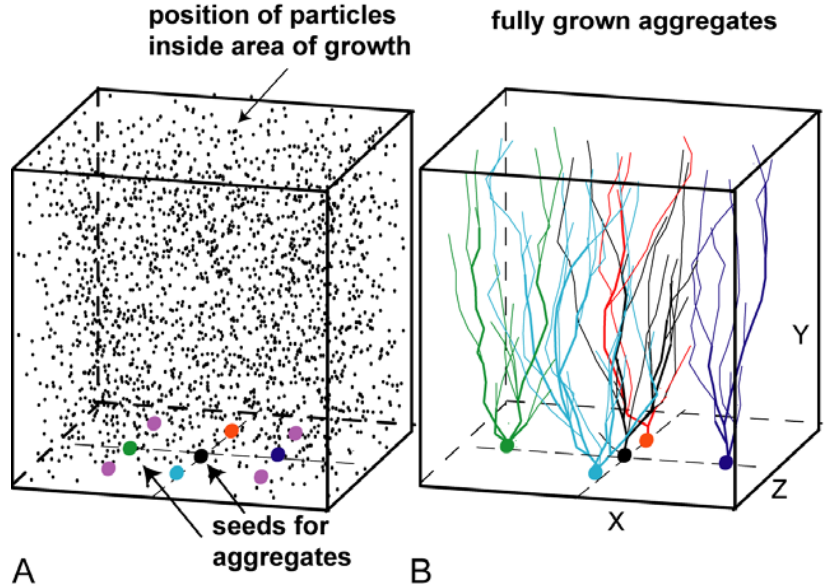

**Figure 2.**
Generating neurons in ensemble. (A) Illustration of the initial condition for generating nine aggregates. (B) Generated granule cells (cells in corners are not shown for visualization clarity). Rectangular box represents a space limitation imposed on the growth of aggregates.

*2.3. Description of the neuronal shape*
    I used the following measures to quantify the geometry of dendritic trees:

    *Lengths ratio* – This is the total length of terminal segments divided by the total length of intermediate segments. I used the dimensionless *Lengths ratio* instead of specifying the total length of dendritic segments, for example, since a calculation of length in microns would depend on the use of a semi-arbitrary scaling factor for the neuron size in my model.

    *PC2/PC1* – This approximates the width/height ratio of a dendrite. Assuming that each point of a neuron is a row vector, I applied principal component analysis to find the two main axes of a dendrite. Here, *PC1* is a SD of the first principal component scores, and *PC2* is a SD of the second principal component scores. For calculating *PC2/PC2, SDD* and "fractal" dimension (see below) the Z-coordinates were set to zero, which decreased the within-group variability and improved the discrimination between neuronal types (the reconstructed neurons have considerably larger shape distortion in Z plane caused by the largest shrinkage of brain slices in that plane; Pyapali and Turner, 1996).

    *Skewness of distances distribution* (*SDD*) – The SDD is a measure of the distribution of pairwise distances between points on the surface of a 3D object, which can provide a scale- and orientation-independent signature of the 3D structure of that object (Osada et al., 2002). The skewness of a distribution is a measure of the asymmetry of the data around the sample mean, and is defined as $SDD = E((x-\mu)^3) / \sigma^3$, where $\mu$ is the mean of $x$ (here $x$ represents the pairwise distances), $\sigma$ is the standard deviation of $x$, and $E(t)$ represents the expected (mean) value of the quantity $t$. For neurons in this study, the distribution of all pairwise distances between the terminal points of each dendrite was calculated (for neurons with more then 50 terminals, 50 randomly selected end points were used, which reliably represented the entire distribution of end points). The skewness of such a distribution provides information about the regularity of the shape. For example, a negative skewness indicates that the data are spread out more to the left of the mean than to the right, meaning that there is a greater proportion of shorter pairwise distances.

**Table I.**
The parameters used to generate different types of neurons. The box size defines the space containing nine growing aggregates (the spatial orientation of the *X*-, *Y*-, and *Z*-axes is illustrated in Fig. 2). The particle density denotes a probability that a given cell (of size 1×1×1) in the box is occupied by an NP. Note that the particle density can change along the *Y* axis, which can correspond to differences between cortical layers.

|  | Box size YxXxZ | Pruning span | Particles density |
|---|---|---|---|
| Granule cell | 20x24x24 | 34 | 0.3 for $0 < Y \leq 4$<br>0.7 for $4 < Y \leq 20$ |
| Basal dendr. | 10x40x40 | 30 | 0.7 for $0 < Y \leq 10$ |
| Apical dendr. | 72x28x28 | 25 | 0.4 for $0 < Y \leq 14$<br>0.2 for $14 < Y \leq 72$ |
| Interneur. axon tree | 10x54x46 | 14 | 0.05 for $0 < Y \leq 7$<br>0.2 for $7 < Y \leq 10$ |
| Purkinje cell | 32x130x14 | 21 | 0.9 for $0 < Y \leq 32$ |

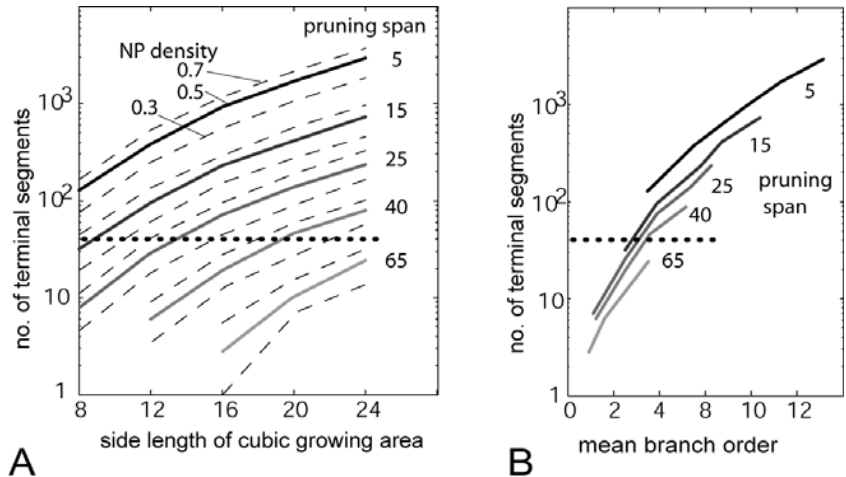

**Figure 3.**
The number of terminal segments of a dendritic tree as a function of model parameters. (A) The number of terminal segments increases with the size of the cubic box, which limits the growth of an aggregate. Increasing the time span of pruning has the opposite effect, reducing the number of ends. The solid lines represent aggregates generated with different pruning span values (numbers on the right) and with the density of NPs set to 0.5. The dashed lines above and below the solid lines represent aggregates generated with NP densities of 0.7 and 0.3, respectively. The horizontal dotted line illustrates that a dendrite with the same number of terminal segments can be generated with different combinations of pruning span values and side lengths. (B) Data points with an NP density of 0.5 from panel A redrawn as a function of mean branch order (for easier comparison between plots, the intensity of the gray color denotes dendrites generated with different pruning span values).

*Asymmetry index* (*A*) – This is a topological measure of a tree based on the number and connectivity pattern of the segments while disregarding the length of segments, their diameters, and the spatial embedding. The asymmetry index is defined as the mean value of the asymmetry of its partitions (subtrees): $A = (n-1)^{-1}\Sigma A_p(r_i,s_i)$. The summation runs over all $n-1$ branch points of the tree with degree $n$, while the partition $(r_i,s_i)$ denotes the number of terminal segments on both subtrees at branch point $i$, and $A_p$ denotes the partition asymmetry: $A_p = |r - s| / (r + s - 2)$, for $r + s > 2$ and $A_p(1,1) = 0$. The asymmetry index ranges from zero for perfectly symmetrical trees to one for perfectly asymmetrical trees (Van Pelt et al.1992).

*Branch order* – The branch order represents the topological distance from the soma. Its value is an integer that is incremented at every bifurcation. A branch order equal to zero is assigned to the primary segments; i.e., those emerging directly from the soma. The maximum branch order and the mean branch order were calculated for every tree in this study.

*"Fractal" dimension* (α,β) – Fractal dimension (FD) is a measure of the degree of object complexity based on how fast measurements increase or decrease as a scale becomes larger or smaller. The fractal dimension of object *S* can be defined as: $FD = -\lim_{e\to 0} \log(N_e) / \log(e)$, where $N_e$ is the minimum number of cubes of side length *e* needed to cover *S*. The neuron however shows a continuous variation of the gradient in the log ($N_e$) / log(*e*) relation with no characteristic slope, and hence it does not properly speaking have a single fractal dimension (Caserta et al., 1995; Smith et al., 1996; Cannon et al., 1999). Therefore, here the fractal properties of neurons were assessed with quantities derived from the caliper method as proposed by Cannon et al. (1999). The caliper method, consists of measuring the apparent length L(λ) when the structure is viewed at various resolutions, defined as different values for λ for the shortest resolvable section. In practice, this amounts to measuring off sections as if with calipers and ignoring features smaller then λ. For a fractal, these quantities show a power law relation: $L(\lambda) \sim \lambda^{1-f}$, where the quantity f in the exponent is termed fractal dimension. As discussed in Cannon et al. (1999), this is not the case for most neurons. They do, however, follow a relation of the form: $\log(L) = \log(L_0) - \exp(\alpha (\log(\lambda) - \beta))$, where $L_0$ is the measured length with the smallest step size. The quantities α and β were found by least squares fitting and have been used here to characterize the "fractal" dimension of neurons. Unlike the real fractal dimension, the parameters α and β do change with the scaling size of an object. Therefore, to calculate α and β the generated neurons were scaled to match the mean branch length of actual neurons. The scaling factor was the following: for basal dendrites – 33, apical dendrites – 15, granule cell – 25, interneurons – 40, Purkinje cells – 6; for the size of NP set to 1.

*2.4. Real neuron morphology data*
Files with intracellularly labeled, reconstructed, and digitalized neurons were obtained from the Duke-Southampton on-line archive of neuronal morphology (http://neuron.duke.edu/cells/cellArchive.html; Cannon et al., 1998). For this analysis, the following groups of neurons were used: 38 granule cells from rat dentate gyrus (unpublished data: Turner and Buzsáki, 1998), 55 CA1 hippocampal pyramidal cells stained with biocytin in whole anesthetized rats (Pyapali and Turner, 1994; Pyapali and Turner, 1996, Pyapali et al., 1998, Turner et al., 1995), 13 interneurons from rat dentate gyrus in brain slices stained with biocytin (Mott et al., 1997), and 3 Purkinje cells from the cerebellar cortex of adult guinea pigs, labeled with horseradish peroxidase, and completely reconstructed from serial sections (Rapp et al., 1994; downloaded from http://www.krasnow.gmu.edu/ascoli/CNG, Ascoli et al., 2001).

## 3. Results

DLA is a model for the formation of fractal-like structures, and hence the choice of the size of the growing area and the corresponding pruning span could be regarded as a choice of resolution (scale) for generating a given type of aggregate. The selection of a pruning span depends on the size of NPs relative to the size of the growing area: smaller NPs produce an aggregate with a finer structure and, as a result, more branches have to be removed to obtain tree with a similar size. This is illustrated in Figure 3A, which shows the number of terminal segments as a function of model parameters. As expected, trees with a larger number of terminals can be produced by: decreasing the pruning span (fewer terminal segments removed), increasing the size of the growing area or increasing the NP density. Interestingly, a tree with a given number of terminals could be generated with different combinations of pruning span value and size of the growing area. For example, a tree with ~40 terminals can be generated with a pruning span value of 15, 25 or 40 and side lengths of the cubic growing space of 9, 13, or 19, respectively (Fig. 3A, dotted line; the above values were obtained for simulations with one seed and with the NP size set to one). Trees with the same number of terminals and generated in a box with the same ratio of side lengths had also similar topological structure despite using different pruning span values. This is illustrated in Figure 3B, where the mean branch order is relatively constant for a given number of terminals irrespective of the pruning span and scaling factor of the growing area, although a systematic shift toward higher branch orders is visible for larger pruning span values. During growth of the aggregate, pruning most affects older branches, which do not have access to new NPs and are thus incapable of regenerating. Such branches are shielded from NPs by newly growing branches. Therefore, pruning during aggregate growth is comparable to deleting branches from a fully grown aggregate. For example, similar dendritic trees could be obtained when only in the final step of the procedure the tree is pruned by recursively deleting terminal branches.

The shape of generated dendrites depends also on the relative side lengths of the rectangular space available for growth and the spatial distribution of NPs (Table I). For example, increasing the height/width ratio of the box changes the dendrite shape from that of a basal dendrite to a granule cell and, ultimately, to an apical dendrite. Decreasing relative width only in the Z-coordinate direction changes it from a basal dendrite to a Purkinje cell (Figs. 4 and 5). Increasing the concentration of NPs increases the density of branches. Changing the spatial distribution of the concentration of NPs influences both the orientation and density of branches. For example, increasing the concentration of NPs in the upper 30% section of the box, while reducing it almost to zero elsewhere, produces an aggregate with the appearance of an interneuron rather than of a basal dendrite (Fig. 5B). Such changes in particle density along the *Y*-axis may be biologically justified as reflecting different cortical layers. In the model, the initial distribution of particle densities along the *Y*-axis exhibits a sharp transition between two regions with different concentrations. However, after a few iterations the diffusive motion of NPs creates a smooth concentration gradient between the layers, which is closer to real biological conditions. Thus, by changing only the space available for growth, the threshold and the spatial distribution of NPs, the DLA model makes it possible to generate 3D structures similar to different types of dendritic and axonal trees (Figs. 4 and 5, Table I).

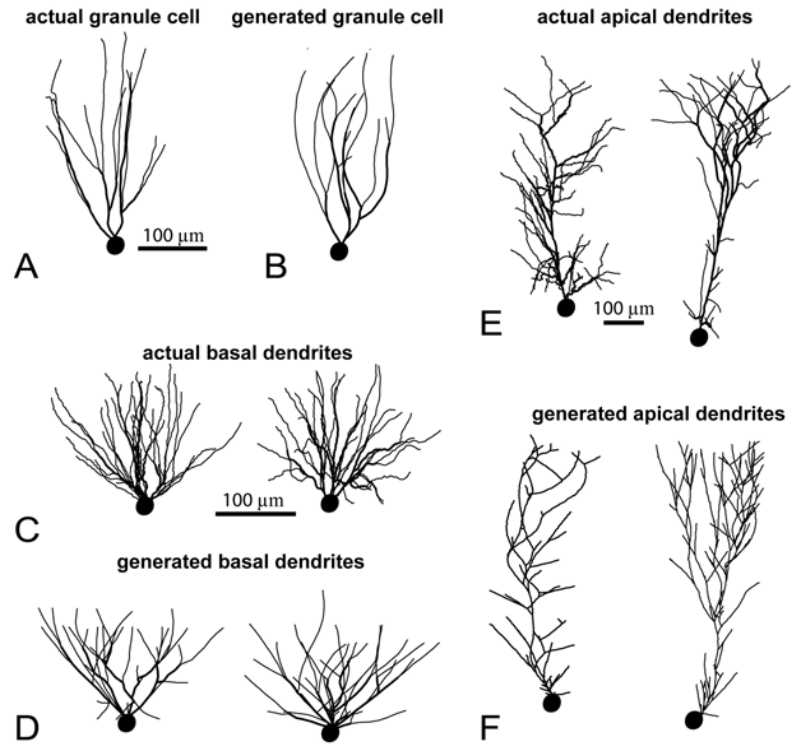

**Figure 4.**
Examples of real and generated neurons. (A, B) Examples of real and generated granule cells. (C, D) Examples of real and generated basal dendrites. (E, F) Examples of real and generated apical dendrites of pyramidal cells. The cell bodies are depicted by spheres.

Quantitatively comparing the generated and real dendrites is challenging because there is no complete measure for describing the complex 3D geometry of a tree. Here, to quantify the geometrical properties of dendritic trees, I calculated the mean and SD of the eight measures described in the Materials and Methods section: the number of terminal branches, the lengths ratio, mean and maximum branch order, asymmetry index, $PC2/PC1$, $SDD$, and "fractal" dimension. The results of a quantitative comparison between generated dendrites (50 dendrites of each type of cell) and real dendrites (55 pyramidal cells, 13 interneurons, 38 granule cells, and 3 Purkinje cells) are summarized in Table II, and examples of generated and real neurons are illustrated in Figures 4 and 5. In almost all cases the mean values of the dendritic geometry measures for the generated neurons were within one SD of the mean values for real neurons (Table II). Note that values calculated for real Purkinje cells may not be very accurate as only three reconstructed cells were available.

The use of a rectangular box to limit neuron growth may, at first sight, appear to impose an artificial constraint, whereas this actually simulates the space limitations imposed by, for example, the extent of the cortex layer, the extent of the area with neurotrophic substances, and by neighboring neurons growing simultaneously (Devries and Baylor, 1997). The present model tested the impact of the last-mentioned factor by generating aggregates in ensemble (9 simultaneously growing cells). In that case, the neighboring aggregates competed for available space and access to NPs, which limited the sideways growth of neurons. The distances between aggregates in this model do not reflect the distances between real neurons, which are actually much closer in the cortex. This is due to the particles constituting the aggregates being relatively large, which reduces the probability that the branches of one aggregate would penetrate space occupied by another aggregate tree. To model a slab of tissue containing a realistic density of cells, the particle size used here would have to be reduced by at least an order of magnitude, which would make the computations prohibitively long (see Materials and Methods). Also due to computation time the number of generated neurons was limited to nine. Nevertheless the model can be easily extended to generate larger number of neurons by increasing the number of seeds and adequately increasing the size of the box in X and Z direction. Given the small amount of cells, most of those cells grew next to the side of the box. In some cases it caused shape distortions when branches grew along the box side.

Competition among aggregates increases when the available space becomes smaller. For example, decreasing the distances between cells results in larger aggregates tending to suppress smaller aggregates by gathering more NPs, which can lead to a drastic 'bigger gets bigger' scenario. This can be prevented by imposing a maximum allowed size for neurons generated in an ensemble, which suggests that an intrinsic limitation on the size of dendritic trees plays an important role in shaping neuron geometry and preventing a winner-take-all space outcome. In the model, this effect can also be reduced by decreasing the size of NPs, thus decreasing the probability that NPs will be caught by a larger aggregate. The simulations also revealed that placing seeds of aggregates at similar depths reduced the differences between the geometries of the dendrites (data not shown). Thus suggesting that lamination of the cortex can facilitate the generation of dendrites with reproducible shapes.

*Trees*
In this paper, I present a simple diffusion model to reproduce spatial embedding of neurons. The same model can also be applied to model diverse types of tree structures. As an example, a generated root and two types of real trees (pear tree and hornbeam) are shown in Figure 6.

**Table II.**
Comparison of geometrical measures (mean and SD values) between generated and real dendritic trees (for interneurons, values were calculated for an axonal tree). The Materials and Methods section provides a description of the measures.

|  | # of ends | Lengths ratio | Branch order | | Asm. index | PC2/PC1 | SDD | Fractal | |
|---|---|---|---|---|---|---|---|---|---|
|  |  |  | mean | max |  |  |  | α | β |
| | | | | Actual neurons | | | | | |
| Granule | 17.1 | 2.79 | 4.29 | 6.84 | 0.45 | 0.66 | 0.39 | 1.06 | 5.61 |
| SD | 4.4 | 1.13 | 0.54 | 1.06 | 0.25 | 0.13 | 0.18 | 0.18 | 0.21 |
| Basal | 34.2 | 4.23 | 5.82 | 9.87 | 0.48 | 0.73 | 0.36 | 1.24 | 4.80 |
| SD | 11.9 | 1.37 | 1.03 | 2.05 | 0.16 | 0.15 | 0.34 | 0.28 | 0.22 |
| Apical | 67.8 | 2.72 | 12.66 | 24.41 | 0.62 | 0.33 | 0.68 | 0.81 | 5.07 |
| SD | 31.9 | 0.74 | 2.77 | 5.85 | 0.09 | 0.08 | 0.19 | 0.14 | 0.22 |
| Intern. | 53.0 | 1.75 | 8.74 | 15.53 | 0.66 | 0.42 | 0.70 | 1.01 | 4.76 |
| SD | 36.3 | 0.25 | 2.50 | 5.22 | 0.13 | 0.16 | 0.12 | 0.23 | 0.28 |
| Purkin. | 437 | 1.07 | 14.70 | 27.00 | 0.52 | 0.78 | 0.25 | 0.71 | 3.55 |
| SD | 31.2 | 0.21 | 0.79 | 2.64 | 0.01 | 0.20 | 0.22 | 0.02 | 0.05 |
| | | | | Generated neurons | | | | | |
| Granule | 17.5 | 2.59 | 4.25 | 7.17 | 0.56 | 0.57 | 0.37 | 1.30 | 5.47 |
| SD | 5.3 | 1.26 | 0.62 | 1.36 | 0.22 | 0.13 | 0.25 | 0.31 | 0.11 |
| Basal | 32.8 | 3.21 | 5.27 | 9.33 | 0.53 | 0.75 | 0.32 | 1.33 | 4.75 |
| SD | 3.3 | 0.52 | 0.46 | 1.05 | 0.16 | 0.07 | 0.11 | 0.21 | 0.06 |
| Apical | 66.7 | 1.47 | 11.24 | 21.04 | 0.71 | 0.25 | 0.65 | 0.85 | 5.01 |
| SD | 42.6 | 0.43 | 4.35 | 8.28 | 0.11 | 0.11 | 0.19 | 0.13 | 0.16 |
| Intern. | 57.9 | 1.74 | 8.03 | 14.12 | 0.61 | 0.45 | 0.50 | 1.06 | 5.04 |
| SD | 32.95 | 0.50 | 2.38 | 4.40 | 0.19 | 0.12 | 0.27 | 0.12 | 0.16 |
| Purkin. | 457 | 1.52 | 14.93 | 28.80 | 0.66 | 0.76 | 0.29 | 0.88 | 3.98 |
| SD | 222 | 0.4 | 3.08 | 7.01 | 0.04 | 0.14 | 0.21 | 0.08 | 0.21 |

## 4. Discussion

The main objective of this work is to illustrate that the creation of complex reproducible dendritic trees does not require precise guidance or an intrinsic plan of the neuron geometry, but rather that external factors can account for the spatial embedding of the major types of dendrites observed in the cortex. In this model the number of terminal branches, the mean and maximum branch orders, and the fractal dimension and other parameters of dendrite geometry are all controlled by a few basic environmental factors. The most important factor in determining the shape of generated neurons is the space available for growth. Changes in the other factors such as the concentration or size of NPs can lead to a similar dendritic shape by adjusting the pruning span of terminals (Fig. 3). In summary, the presented DLA-based model reveals that a simple, diffusive growth mechanism is capable of creating complex and diverse 3D trees strictly similar to observed neuronal shapes.

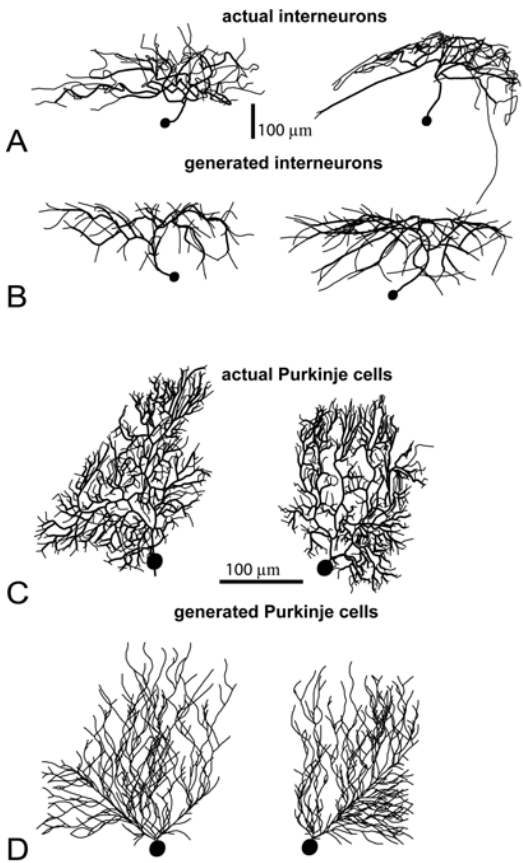

**Figure 5.**
Examples of real and generated neurons. (A, B) Examples of real and generated axonal trees of interneurons. (C, D) Examples of real and generated Purkinje cells. The cell bodies are depicted by spheres.

In the DLA model, connecting a new particle to the aggregate approximates growth in the direction of a local gradient. DLA is similar to Laplacian growth where the probability of growth at any point on the boundary of the growing object is determined by Laplace's equation, which describes the 'attraction' field around the object (Hastings and Levitov, 1998). Therefore the growth in the direction of a local gradient and the DLA model incorporating connecting particles to the aggregate are almost equivalent. Thus I have used DLA as a computationally convenient tool to model (1) the growth of a dendrite toward a higher concentrations of NPs, (2) diffusive motion of NPs, and (3) competition between dendrites for access to NPs.

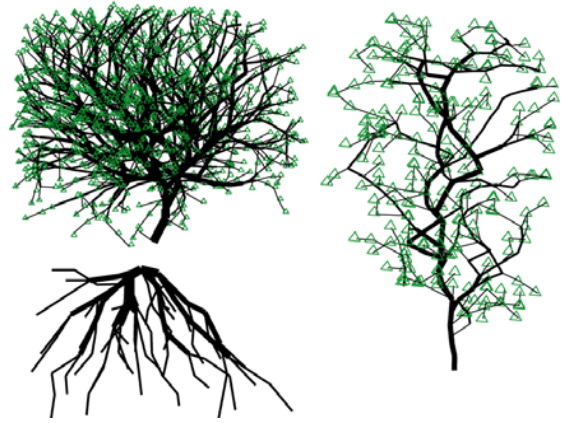

**Figure 6.**
Demonstration of the general applicability of the algorithm to model diverse types of tree structures (from left: pear tree, root and hornbeam; terminal branches are depicted as triangles to resemble leaves).

The real dendrites grow by elongation and can branch either via bifurcation of growth cone-like tips or through interstitial sprouting of new branches from an existing dendritic branch. These new branches extend and retract to undergo constant remodeling. Only a subset is eventually stabilized (Jan and Jan, 2003). This phenomenon of constant pruning of dendritic branches during neuron development is modeled here by probabilistic deleting the terminals. Parts of neuron, which were not deleted during a specified number of iteration (pruning span), become "stabilized" by being excluded from any further pruning. The growth and pruning of real cortical neurons is strongly influenced by excess or deficit of extrinsic factors, which includes for example: neurotrophin 3, brain-derived neurotrophic factor (BDNF) and nerve growth factor (McAlister et al., 1997). For instance, BDNF released from an individual cell alters the structure of nearby dendrites on an exquisitely local scale (Horch and Katz, 2002). The intrinsic factors have an effect on stability rather than directionality of the dendrite by affecting the dynamics of structural components of dendrites (Scott and Luo, 2001). The NPs in the presented model do not refer to any concrete neurotrophic substance. I chose to call those particles 'neurotrophic' to suggest a biological interpretation of the model, which is that, a new dendrite branch sprouts at the point of contact with neurotrophic particles. Stated differently, connecting NP to the aggregate can be seen as equivalent to the process where a new part of a dendrite came from the cell itself at the location where the NP was detected. Also, a decrease in the number of freely moving NPs after contacting the aggregate has a biological justification, namely that the neurotrophic molecules are commonly uptaken by neurons and transported to the cell body (Purves, 1988; von Bartheld et al., 1996). As mentioned above the neurons' development is a very complicated process and the model presented here cannot account for all possible phenomena affecting neurons shape. For example, the morphology of axons and dendrites can be affected by mechanical tensions during brain development (Van Essen, 1997). Additional model parameters could improve the model's accuracy, but would also increase its complexity. Thus, in light of the fact that the existing model performs well and the goal of keeping the model simple, I believe the model's current level of complexity and accuracy are appropriately balanced.

It is notable that the presented model uses only five or seven parameters (depending on the number of layers) to reproduce complex and diverse neuronal shapes: three dimensions characterizing space, a pruning span, and one or three parameters to specify concentrations of NPs depending on whether one or two layers are considered, respectively. For comparison, van Pelt et al. (1997) uses three free parameters in his one-dimensional BES-model of neuronal growth, and several more parameters were used by Samsonovich and Ascoli (2005) in their 3D model of hippocampal cells.

Besides investigating the role of environmental factors in shaping dendritic geometry, the presented model can also be of benefit for modeling community. The ever-increasing computational power of computers allows more realistic models of the cortex to be considered, which include connectivity patterns between neurons, their electrophysiological properties, and full dendritic and axonal geometry (Muhammad and Markram, 2005; Ascoli, 1999). This type of realistic large-scale modeling requires at least hundreds of neurons. Due to the lack of such a large number of fully reconstructed different types of neurons, these models may benefit from the use of synthetic cells. The model developed in this study can provide a new means for generating a large number of synthetic neurons. The software written by the author to generate the presented types of neurons is freely available.

*Conclusions*
In this paper I have proposed a single mechanism for the formation of diverse neuron shapes. The results demonstrate that simultaneously grown diffusion-limited aggregates competing for available resources create reproducible self-organized structures that are strikingly similar to neurons (Figs. 4 and 5). This is the first model to simulate 3D neuronal growth accounting for external factors such as the NP concentration, competition between neurons, and space limitations. Moreover, it advances DLA-based models by incorporating pruning and space limitations. Analysis of the discrepancies between generated and real neurons may elucidate the relative contribution of other factors that – together with environmental factors – affect neuron outgrowth. Finally, the presented model is readily applicable to the modeling and analyses of other types of tree structures, as shown by the example given in Figure 6.

**Acknowledgments**
I thank the anonymous reviewers for their helpful comments and suggested improvements. I also thank L. Pelechacz for helping to edit the initial version of the manuscript.